\documentclass{emulateapj}
\usepackage{natbib}

\usepackage{amssymb}
\usepackage{amsthm}
\usepackage[figuresright]{rotating}

\newcommand{\al}{$\alpha$}
\newcommand{\g}{$\gamma$}
\newcommand{\raa}{($\alpha$,$\alpha$)}
\newcommand{\raX}{($\alpha$,$X$)}

\newcommand{\rag}{($\alpha$,$\gamma$)}
\newcommand{\ran}{($\alpha$,n)}

\newcommand{\rap}{($\alpha$,p)}

\newcommand{\stot}{$\sigma_{\mathrm{reac}}$}

\newcommand{\Nsv}{$N_A$$\left< \sigma v \right>$}

\newcommand{\zrvi}{$^{96}$Zr}

\newcommand{\moix}{$^{99}$Mo}

\begin{document}

\title{Low Energy measurement of the $^{96}\mathrm{Zr}(\alpha,n)^{99}\mathrm{Mo}$ reaction cross section and its impact on weak r-process nucleosynthesis}

\author{G.\,G.\,Kiss}
\affil{Institute for Nuclear Research (ATOMKI), H-4026 Debrecen, Bem t\'er 18/c, Hungary}
\email{ggkiss@atomki.hu}

\author{T.\, N.\,Szegedi}
\affil{Institute for Nuclear Research (ATOMKI), H-4026 Debrecen, Bem t\'er 18/c, Hungary\\
University of Debrecen, H-4001 Debrecen, Egyetem t\'er 1, Hungary}

\author{P.\,Mohr}
\affil{Institute for Nuclear Research (ATOMKI), H-4026 Debrecen, Bem t\'er 18/c, Hungary}

\author{M.\ Jacobi}
\affil{Institut f{\"ur} Kernphysik, Technische Universit{\"a}t
  Darmstadt, Schlossgartenstr.~2, D-64289 Darmstadt, Germany}

\author{Gy.\,Gy\"urky}
\affil{Institute for Nuclear Research (ATOMKI), H-4026 Debrecen, Bem t\'er 18/c, Hungary}

\author{R.\,Husz\'ank}
\affil{Institute for Nuclear Research (ATOMKI), H-4026 Debrecen, Bem t\'er 18/c, Hungary}

\author{A.\ Arcones}
\affil{Institut f{\"ur} Kernphysik, Technische Universit{\"a}t
  Darmstadt, Schlossgartenstr.~2, D-64289 Darmstadt, Germany\\
GSI Helmholtzzentrum für Schwerionenforschung GmbH,
  Planckstr. 1, D-64291 Darmstadt, Germany \\
Helmholtz Forschungsakademie Hessen f\"ur FAIR, GSI Helmholtzzentrum für
Schwerionenforschung, 64291 Darmstadt, Germany}

\begin{abstract}
Lighter heavy elements beyond iron and up to around silver can form in neutrino-driven ejecta in core-collapse supernovae and neutron star mergers. Slightly neutron-rich conditions favour a weak r-process that follows a path close to stability. Therefore, the beta decays are slow compared to the expansion time scales, and ($\alpha$,n) reactions become critical to move matter towards heavier nuclei. The rates of these reactions are calculated with the statistical model and their main uncertainty, at  energies relevant for the weak r-process, is the $\alpha$+nucleus optical potential. There are several sets of parameters to calculate the $\alpha$+nucleus optical potential leading to large deviations for the reaction rates, exceeding even one order of magnitude. Recently the $^{96}$Zr($\alpha$,n)$^{99}$Mo reaction has been identified as a key reaction that impacts the production of elements from Ru to Cd. Here, we present the first cross section measurement of this reaction at energies (6.22 MeV $\leq$ E$_\mathrm{c.m.}$ $\leq$ 12.47 MeV) relevant for the weak r-process. The new data provide a stringent test of various model predictions which is necessary to improve the precision of the weak r-process network calculations. The strongly reduced reaction rate uncertainty leads to very well-constrained nucleosynthesis yields for $Z = 44 - 48$ isotopes under different neutrino-driven wind conditions.

\end{abstract}

\section{Introduction}
\label{sec:int}

Half of the stable isotopes heavier than iron are produced by the rapid neutron capture process (r-process) when neutron captures are faster than beta decays. This process requires extreme neutron densities and explosive environments, therefore the two favourite candidates are: core-collapse supernovae, where neutron stars are born, and neutron star mergers. After a successful core-collapse supernova, there is a neutrino-driven wind  consisting of matter ejected by neutrinos emitted from the hot proto-neutron star. For many years, this was the preferred scenario for the r-process, even if the conditions were only  slightly neutron rich or proton rich and thus not enough for the r-process (for a review see \cite{Arcones2013a} and reference therein). In contrast, the r-process has been observed in neutron star mergers. After the gravitational wave detection of GW170817 \citep{Abbott2017a}, there was an observation of the kilonova light curve produced by the radioactive decay of the neutron-rich nuclei formed during the r-process \citep{Metzger2010a, Abbott2017a}. Also Sr was directly observed in the kilonova spectrum \citep{Watson2019a}. Still there are many open questions concerning the astrophysical site and the nuclear physics involved.

Observations of the oldest stars in our galaxy and in neighbour dwarf galaxies \cite[see e.g.,][]{Frebel2018a, Reichert2020a, Cote2019a} indicate that the r-process occurred already very early, even before neutron star mergers could significantly contribute. This points to  rare supernovae, and recent investigations have shown that magneto-rotational supernovae could account for this early r-process contribution \cite[see e.g.,][]{Winteler2012a, Nishimura2017a, Moesta2018a, Reichert2020b}. Another hint from observations is that the elements between Sr and Ag may be produced by a separate or additional process to the r-process \citep{Travaglio2004a, Qian2000a, Montes2007a, Hansen2014a}. One possibility to explain these observations is the neutrino-driven ejecta from core-collapse supernovae \citep{Qian1996a, Wanajo2011a, Arcones2011a, Arcones2014a}.

In neutrino-driven, neutron-rich supernova ejecta, the weak r-process can form the lighter heavy elements between Sr and Ag \cite[see e.g.,][]{Bliss2018a}. Initially the matter is close to the neutron star and very hot, therefore a nuclear statistical equilibrium (NSE) is established. As matter expands and cools down, individual nuclear reactions become important at temperatures below about 5 GK. \cite{Bliss2018a} have investigated all possible conditions expected in neutrino-driven, neutron-rich supernova ejecta and identified those where nuclear reactions are important. In the weak r-process, the nucleosynthesis path is determined by (n,$\gamma$)-($\gamma$,n) equilibrium and stays close to stability. Consequently, compared to the expansion timescale, at temperatures above about 2 GK, $\beta$ decays are too slow to move matter to higher proton numbers and ($\alpha$,n) and (p,n) reactions become faster \citep{Bliss2017a}.

Therefore, in order to use observations to understand the astrophysical conditions where lighter heavy elements are produced, one has to reduce the nuclear physics uncertainties of the key reactions. In a broad sensitivity study \citep{Bliss2020a}, several ($\alpha$,n) reactions have been identified as critical because of their impact on the abundances under different astrophysical conditions. These reactions rates are calculated from the cross sections computed with the Hauser-Feshbach statistical model which relies on nuclear physics inputs. Recently, a series of sensitivity calculations were performed to evaluate the theoretical uncertainty of these cross section calculations \citep{Pereira2016a, Mohr2016a, Bliss2017a}. These works identified different $\alpha$+nucleus optical potential parameter sets ($\alpha$OMP's) as the main source of uncertainty. The difference between the cross section based on  various $\alpha$OMP's can exceed even an order of magnitude \citep{Pereira2016a}. Therefore, experiments are critical to reduce the uncertainties of the rates. Low energy alpha-induced reaction cross section measurements were frequently used to constrain the parameters of the $\alpha$OMP's used in astrophysical calculations \citep{Sauerwein2011a, Scholz2014a, Kiss2015a}. However, such precise experimental data, reaching sub-Coulomb energies are typically missing for isotopes located at or close to the weak r-process path \citep{Bliss2017a}.

Here we contribute to a more reliable weak r-process calculation by measuring the $^{96}$Zr($\alpha$,n)$^{99}$Mo reaction cross section for the first time at energies relevant for the weak r-process nucleosynthesis and by using the precise data to evaluate the $\alpha$OMP's used in the nucleosynthesis network. This reaction is one of the bottlenecks that sensitively affects the production of nuclei between 44 $\leq$ Z $\leq$ 47 \citep{Bliss2020a}. We demonstrate that reducing the nuclear physics uncertainty to a 30\% level is critical and enough to get accurate abundance predictions.

\begin{figure*}
\center
\resizebox{0.85\textwidth}{!}{\rotatebox{0}{\includegraphics[clip=]{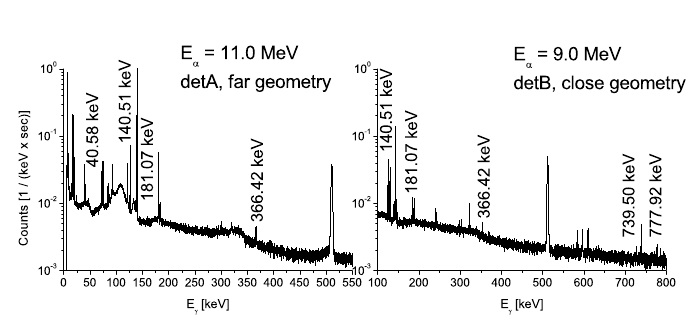}}}
\caption{\label{fig:spectra}
$\gamma$-ray spectra, measured for one hour, taken on detA, t$_{waiting}$ = 9.5 h after the 11 MeV irradiation (left panel); and on detB, t$_{waiting}$ = 62.6 h after the 9 MeV irradiation (right panel). The peaks used for the analysis are marked.
}
\end{figure*}

This paper is structured as follows. In Sect.~\ref{sec:exp}, we present our experimental approach. The results including a theoretical analysis are in Sect.~\ref{sec:res}, and the impact of those on the weak r-process is in Sect.~\ref{sec:rproc}. Finally, we provide a short summary and conclusions are given in Sect.~\ref{sec:sum}.

\section{Experimental approach}
\label{sec:exp}

The cross section measurement was carried out at the Institute for Nuclear Research (Atomki) 
using the activation technique. The targets were prepared by electron beam evaporation of metallic Zr onto 6 $\mu$m thick Al foil backing. Similarly to our previous cross section measurements \citep{Korkulu2018a, Kiss2018a}, the absolute number of target atoms was determined with the Rutherford Backscattering technique using the Oxford-type Nuclear Microprobe Facility at Atomki \citep{Huszank2016a}. The energy and the diameter of the beamspot of the $^4$He$^+$ beam provided by the Van de Graff accelerator was 2.0 MeV  and 2.5 $\mu$m, respectively. Two Silicon ion-implanted detectors (50 mm$^2$ sensitive area and 18 keV energy resolution) were used to measure the yield of the backscattered ions, one of them was placed at a scattering angle of 165$^{\circ}$ and the other one was set to 135$^{\circ}$. Target thicknesses between 1.23 x 10$^{18}$ and 1.54 x 10$^{18}$ Zr atom/cm$^2$ were found with an uncertainty of typically 5\%. This total uncertainty was derived as the quadratic sum of the following uncertainties: measurements of the RBS standards (3\%), statistical uncertainty ($\leq$ 3\%) and uncertainty of the isotopic abundance (3.2\%).”

\begin{table}
\center
\caption{\label{tab:decay}Decay parameters of the reaction product $^{99}$Mo and its daughter $^{99}$Tc$^m$, taken from \cite{Browne2017a, Goswamy1992a}.}
\begin{tabular}{ccccc}
\hline
\multicolumn{1}{c}{Residual} &
\multicolumn{1}{c}{Half-} &
\multicolumn{1}{c}{Energy} &
\multicolumn{1}{c}{Relative } \\
\multicolumn{1}{c}{nucleus} &
\multicolumn{1}{c}{life [h]} &
\multicolumn{1}{c}{[keV]} &
\multicolumn{1}{c}{intensity [\%]} \\
\hline
$^{99}$Mo    &  65.924 $\pm$ 0.006   & 40.58 &  1.04 $\pm$ 0.03 \\
&& 181.07 & 6.05 $\pm$ 0.12 \\
&& 366.42 & 1.20 $\pm$ 0.02 \\
&& 739.50 & 12.20$\pm$ 0.02 \\
&& 777.92 & 4.31 $\pm$ 0.08 \\
$^{99}$Tc$^m$  &  6.0072 $\pm$ 0.0009 & 140.51 & 89 $\pm$ 4\\
\hline
\end{tabular}
\end{table}

The Zr targets were irradiated with $\alpha$ beams from the MGC cyclotron of Atomki. The energy of the $\alpha$ beam was between $E_\mathrm{lab}$ = 6.5 MeV and $E_\mathrm{lab}$ = 13.0 MeV, this energy range was scanned with energy steps of 0.5 MeV - 1.0 MeV. The length of the irradiations varied between $t_\mathrm{irrad}$ = 6 h to $t_\mathrm{irrad}$ = 48 h with beam currents of 0.5 - 1.4 $\mu$A. Longer irradiations were carried out at lower energies to (partially) compensate the lower cross sections. The number of the impinging $\alpha$ particles was obtained from current measurement. After the beam-defining aperture, the chamber was insulated and secondary electron suppression voltage of $-300$~V was applied at the entrance of the chamber. From the last beam-defining aperture the whole chamber served as a Faraday cup. The collected charge was measured with a current integrator, the counts were recorded in multichannel scaling mode, stepping the channel in every minute to take into account the possible changes in the beam current.

The cross sections were measured using the activation technique \citep{Gyuerky2019a}. The decay parameters of the $^{99}$Mo reaction product are summarized in Table \ref{tab:decay}. The $\beta^-$ decay of $^{99}$Mo is followed by the emission of numerous, relatively intense $\gamma$-rays, which were detected by two Germanium detectors: a Low Energy Photon Spectrometer (detA) and a 50\% relative efficiency HPGe detector (detB), both equipped with a 4$\pi$ lead shield. DetA has low laboratory background (about 0.585 1/s in the 50-2000 keV energy region), its resolution is excellent, but with increasing $\gamma$-ray energies its detection efficiency decreases sharply. Accordingly, this detector was used to measure the yield of the $E_{\gamma}$ = 40.58~keV, $E_{\gamma} = 140.51$~keV (belonging to the daughter isotope $^{99}$Tc$^m$), $E_{\gamma} = 181.07$~keV and $E_{\gamma} = 366.42$~keV transitions. The laboratory background of detB is higher (about 5.071~1/s in the 100-2000~keV energy region), however, its detection efficiency is much higher for the higher energy $\gamma$-rays, therefore, this detector was used to measure the yield of the $E_{\gamma} = 739.50$~keV and $E_{\gamma} = 777.92$~keV $\gamma$-rays, also. After the irradiations, $t_\mathrm{waiting}\approx 2.0$~h waiting time was used in order to let the short-lived, disturbing activities decay. The duration of the $\gamma$-countings were two-to-six days in the case of each irradiation and the spectra were saved in every hour. Typical off-line $\gamma$ spectra, measured with detA (left panel) and detB (right panel), can be seen in Fig. \ref{fig:spectra}. The activity of the samples irradiated at $E_\mathrm{lab} = 8$~MeV and higher were measured with both detectors, the resulting cross sections were found to be always consistent. The half-life of the $^{99}$Mo is known from large number of experiments with uncertainty less than 0.01\% \citep{Stone2014a}. The activity of the samples irradiated with alpha beams of $E_\mathrm{lab} = 12$~MeV and $E_\mathrm{lab} = 13$~MeV energies were measured for more than 2 weeks, the deadtime and relative intensity corrected peak areas were fitted with exponential using the least square method. The resulted half-lives, having $\chi^2$ always below 1.3, are in agreement with the literature value within their uncertainties, which proves that no other $\gamma$ transitions pollutes the peaks used for cross section determination.

The low yields measured in the present work necessitated the use of short source-to-detector distances for the $\gamma$-countings carried out after the irradiation of the Zr targets with alpha beams of $E_\mathrm{lab} = 9.0$~MeV and below. The absolute detection efficiency was derived for both detectors using the following procedure: first, using calibrated $^{60}$Co, $^{133}$Ba, $^{137}$Cs, $^{152}$Eu, and $^{241}$Am sources, the absolute detector efficiency was measured in far geometry: at 15 cm and 21 cm distance from the surface of detA and detB, respectively. Since the calibration sources (especially $^{133}$Ba, $^{152}$Eu) emit multiple $\gamma$-radiations from cascade transitions, in close geometry no direct efficiency measurement was carried out. Instead, in the case of the high energy irradiations (at and above 10~MeV) the yield of the investigated $\gamma$-rays was measured both in close and far geometry. Taking into account the time elapsed between the two countings, a conversion factor of the efficiencies between the two geometries could be determined and used henceforward in the analysis.

\section{Results and theoretical analysis}
\label{sec:res}

\begin{figure}
\center
\resizebox{0.99\columnwidth}{!}{\rotatebox{0}{\includegraphics[clip=]{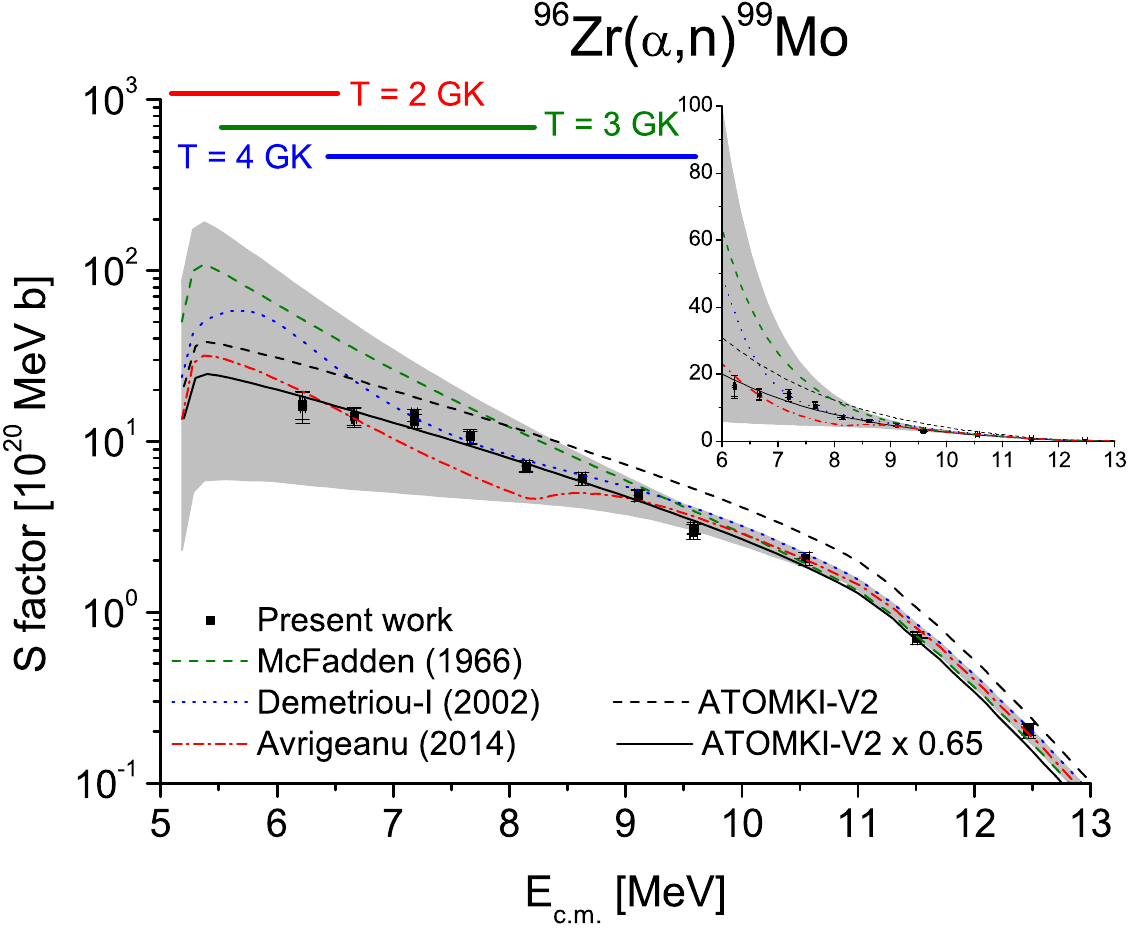}}}
\caption{\label{fig:results} Comparison of experimental and theoretical
  astrophysical S-factors of $^{96}$Zr($\alpha$,n)$^{99}$Mo reaction
  as a function of the energy (the inset shows the same data on linear
  scale). The colored lines indicate the effective Gamow windows for $^{96}$Zr($\alpha$,n)$^{99}$Mo at T = 2 GK, 3 GK, and 4 GK which deviate from the classical
Gamow windows because of the energy dependence of the astrophysical S-factor; see also \cite{Rauscher2010a}. Excellent agreement with $\chi^2/N < 1$ is only obtained for the
  ATOMKI-V2 calculation, scaled by a factor of 0.65 (full line). The wide range of
  TALYS predictions is indicated by the grey-shaded area. Further discussion
  see text.
  }
\end{figure}

The measured $^{96}$Zr($\alpha$,n)$^{99}$Mo cross section values are listed in
Table \ref{tab:res}. The effective center-of-mass energy in the second column
takes into account the energy loss of the beam in the target. The quoted
uncertainty in the $E_\mathrm{c.m.}$ values corresponds to the energy stability of
the $\alpha$-beam and to the uncertainty of the energy loss in the target,
which was calculated using the SRIM code \citep{SRIM}. The activity of several
targets were measured using both detA and detB, in these cases the cross
sections were derived from the averaged results weighted by the statistical
uncertainty of the measured values. The uncertainty of the cross sections is
the quadratic sum of the following partial errors: detection efficiency (5\%),
far-to-close detection efficiency correction factor ($\leq$ 2\%), number of
target atoms (5\%), current measurement (3\%), uncertainty of decay parameters
($\leq$ 4\%) and counting statistics ($\leq$ 15.3\%). The astrophysically relevant energy region (the so-called Gamow window) ranges from E$_{min}$ = 5.1 MeV up to E$_{max}$ = 6.5 MeV at T = 2 GK temperature, from E$_{min}$ = 5.5 MeV up to E$_{max}$ = 8.4 MeV at T = 3 GK temperature and from E$_{min}$ = 6.5 MeV up to E$_{max}$ = 9.6 MeV at T = 4 GK temperature, respectively.
The new data are shown
in Fig.~\ref{fig:results}, and a comparison to theoretical predictions is
made.

The $^{96}$Zr($\alpha$,n)$^{99}$Mo reaction was already studied in several works \citep{Chowdhury1995a, Pupillo2015a, Hagiwara2018a, Murata2019a}. However, because the literature data do not reach the lowest energies, and due to the large c.m. energy uncertainties of the literature data, the theoretical analysis is restricted to our new experimental results.
\begin{table}
\center
\caption{\label{tab:res} Measured cross sections of the $^{96}$Zr($\alpha$,n)$^{99}$Mo reaction. The last five rows show the
average results (weighted by the statistical uncertainties) of the measurements carried out at the same energy.}
\begin{tabular}{cl}
\hline
\multicolumn{1}{c}{$E_\mathrm{c.m.}$} &
\multicolumn{1}{c}{Cross section}\\
\multicolumn{1}{c}{[MeV]} &
\multicolumn{1}{c}{[mbarn]} \\
\hline
	6.22	$\pm$	0.02	&	(2.49	$\pm$	0.52) $\cdot$ 10$^{-4}$ \\
	6.22	$\pm$	0.02	&	(2.54	$\pm$	0.48) $\cdot$ 10$^{-4}$ \\
	6.67	$\pm$	0.02	&	(1.65	$\pm$	0.22) $\cdot$ 10$^{-3}$	\\
	6.66	$\pm$	0.02	&	(1.64	$\pm$	0.19) $\cdot$ 10$^{-3}$	\\
	7.19	$\pm$	0.02	&	(1.32	$\pm$	0.13) $\cdot$ 10$^{-2}$	\\
	7.18	$\pm$	0.02	&	(1.42	$\pm$	0.11) $\cdot$ 10$^{-2}$	\\
	7.66	$\pm$	0.02	&	(6.29	$\pm$	0.62) $\cdot$ 10$^{-2}$	\\
	7.66	$\pm$	0.02	&	(6.29	$\pm$	0.49) $\cdot$ 10$^{-2}$	\\
	8.14	$\pm$	0.03	&	(2.14	$\pm$	0.17) $\cdot$ 10$^{-1}$	\\
	8.62	$\pm$	0.03	&	(7.90	$\pm$	0.67) $\cdot$ 10$^{-1}$	\\
	9.11	$\pm$	0.03	&	(2.48	$\pm$	0.19) $\cdot$ 10$^{0}$ \\
	9.58	$\pm$	0.03	&	(5.27	$\pm$	0.51) $\cdot$ 10$^{0}$ \\
	9.59	$\pm$	0.03	&	(5.58	$\pm$	0.43) $\cdot$ 10$^{0}$ \\
	10.55	$\pm$	0.03	&	(3.48	$\pm$	0.29) $\cdot$ 10$^{1}$	\\
	11.51	$\pm$	0.04	&	(8.38	$\pm$	0.71) $\cdot$ 10$^{1}$	\\
	12.47	$\pm$	0.04	&	(1.33	$\pm$	0.12) $\cdot$ 10$^{2}$	\\
	12.47	$\pm$	0.04	&	(1.36	$\pm$	0.10) $\cdot$ 10$^{2}$	\\
\hline
\hline
	6.22	$\pm$	0.02	&	(2.52	$\pm$	0.48) $\cdot$ 10$^{-4}$	\\
	6.66	$\pm$	0.02	&	(1.64	$\pm$	0.18) $\cdot$ 10$^{-3}$	\\
	7.18	$\pm$	0.02	&	(1.38	$\pm$	0.11) $\cdot$ 10$^{-2}$  \\
	7.66	$\pm$	0.02	&	(6.29	$\pm$	0.49) $\cdot$ 10$^{-2}$  \\
	9.59	$\pm$	0.03	&	(5.47	$\pm$	0.41) $\cdot$ 10$^{0}$	\\
	12.47	$\pm$	0.04	&	(1.35	$\pm$	0.10) $\cdot$ 10$^{2}$	  \\
\hline
\end{tabular}
\end{table}

The new experimental data for the \zrvi \ran \moix\ reaction have been
analyzed in the statistical model (SM). The presented calculations were
performed with the TALYS code. In a schematic notation,
the cross section of an \al -induced \raX\ reaction is given by
\begin{equation}
\sigma(\alpha,X) \sim \frac{T_{\alpha,0} T_X}{\sum_{T_i}} = T_{\alpha,0}
  \times b_X
\label{eq:SM}
\end{equation}
with the transmission coefficients $T_{\alpha,0}$ of the incoming \al -particle, $T_i$ for
the outgoing particles ($i = \gamma, p, n, \alpha, 2n$, etc.), and the
branching ratio $b_X = T_X / \sum_i T_i$ for the branching into the $X$
channel. Usually, the transmissions $T_i$ are calculated from optical model
potentials for the particle channels and from the \g -ray strength function
for the \rag\ capture channel. For further details, see
e.g.~\cite{Rauscher2000a,Rauscher2011a}.

For the \zrvi \ran \moix\ reaction in the energy range under study (see
Fig.~\ref{fig:results}) the neutron channel is dominating
because the proton channel is closed or suppressed by the Coulomb barrier and
the \g -channel is typically much weaker than the neutron channel. Thus, the
branching ratio to the neutron channel is $b_n \approx 1$, and the \ran\ cross
section is almost identical to the total \al -induced reaction cross section
\stot . From Eq.~(\ref{eq:SM}) it can be seen that the \ran\ cross section is
essentially defined by the transmission $T_{\alpha,0}$ which in
turn depends only on the chosen $\alpha$OMP. Other ingredients of the statistical model affect the branching ratios
$b_X$, but have only minor influence on the \ran\ cross section because of
$b_n \approx 1$. For completeness we note that below the \ran\ threshold at
5.1~MeV, we find $b_\gamma \approx 1$, and the \rag\ cross section approaches
the total cross section \stot.

It is obvious from Fig.~\ref{fig:results} that predictions of the \ran\ cross
sections in the SM vary over more than one order of magnitude at the lowest
energies whereas at energies above 10~MeV most predictions agree nicely. For
better readability of Fig.~\ref{fig:results}, we restrict ourselves to the
presentation of the widely used $\alpha$OMP's by 
\cite{McFadden1966a}, 
\cite{Demetriou2002a}, and 
\cite{Avrigeanu2014a}; the latter is the default $\alpha$OMP in TALYS (which is a widely used nuclear reaction code; the calculations shown in Fig.~\ref{fig:results} were carried out using version 1.9).

The reason for the wide range of predictions was identified and discussed in
\cite{Mohr2020a}. The usual SM calculations show a dramatic sensitivity to the
tail of the imaginary potential. To avoid this sensitivity, an alternative
approach was suggested in \cite{Mohr2020a} to use a pure barrier transmission
model (PBTM) for the calculation of the total reaction cross section \stot
. Furthermore, because the PBTM does not allow to predict \ran\ cross
sections, the new ATOMKI-V2 $\alpha$OMP was introduced in the Supplement of
\cite{Mohr2020a}. ATOMKI-V2 consists of the real part of ATOMKI-V1 in
combination with a short-range imaginary part and thus approximates the PBTM
results of \cite{Mohr2020a} for \stot . In addition, ATOMKI-V2 has been
implemented into TALYS; this allows now  the calculation of \ran , \rap , and
\rag\ cross sections within the SM in the usual way which was not possible
within the simple PBTM approach.

The ATOMKI-V2 potential reproduces measured \ran\ cross sections over a wide
range of masses and energies with deviations below a factor of two. This holds
also for the present \zrvi \ran \moix\ reaction (see
Fig.~\ref{fig:results}). However, there is a slight overestimation of the
experimental results over the full energy range under study (dotted line in
Fig.~\ref{fig:results}). Therefore the calculation from the ATOMKI-V2
potential was scaled by a factor of 0.65 to obtain best agreement with the new
experimental data. These scaled cross sections were used to calculate the
astrophysical reaction rate \Nsv\ (see below).

Although the scaling factor of 0.65 is within the estimated uncertainty of the
new approach of \cite{Mohr2020a}, a brief discussion of this factor is
appropriate:

($i$) Technically, the ATOMKI-V2 potential is a complex $\alpha$OMP which approximates the
calculations in the PBTM with small deviations. In the present case, the
ATOMKI-V2 calculation of the total cross section \stot\ is about 10\% higher
than the underlying PBTM calculation.

($ii$) The ATOMKI-V2 potential distinguishes between semi-magic and non-magic
target nuclei; the latter (like \zrvi\ in this work) require a deeper
potential with volume integrals of $J_R = 371$ MeV\,fm$^3$ whereas the
semi-magic targets are characterized by a lower $J_R = 342.4$
MeV\,fm$^3$. Depending on energy, the lower $J_R$ for semi-magic targets
increases the effective barrier and thus
reduces \stot\ by about $15 - 25$\%. An analysis of \zrvi \raa \zrvi\ elastic
scattering at 35 MeV \citep{Lund1995a,Lahanas1986a} requires volume integrals around
$J_R \approx 350$ MeV\,fm$^3$, thus indicating that \zrvi\ behaves more like a
semi-magic nucleus. As a consequence, the usage of the global value $J_R = 371$ MeV\,fm$^3$
instead of the locally optimized $J_R \approx 350$ MeV\,fm$^3$ leads to
an overestimation of \stot\ by about $10 - 20$\%.

Combining the above arguments ($i$) and ($ii$) provides a reasonable
explanation for the obtained scaling factor of 0.65 for the ATOMKI-V2 result
using the global $J_R = 371$ MeV\,fm$^3$ for non-magic target nuclei.

Finally, the agreement of the scaled ATOMKI-V2 calculation with the
experimental data is excellent with $\chi^2$ per point of about 0.6 whereas
calculations with the different $\alpha$OMPs within TALYS show a different energy
dependence (see Fig.~\ref{fig:results}) and cannot reach $\chi^2/N < 3$ (even
with arbitrary scaling factors). The ATOMKI-V2 approach, scaled by the factor
of 0.65, is thus the preferred option for the calculation of the astrophysical
reaction rate \Nsv .

The lowest experimental data point at about 6.2 MeV is located only 1.1 MeV
above the \ran\ threshold at 5.1 MeV. The astrophysical reaction rate
\Nsv\ results from the folding of a Maxwell-Boltzmann velocity distribution
with the energy-dependent cross section $\sigma(E)$. At higher temperatures (at and above 3 GK),
the folding integral is essentially determined by the new experimental
data. At lower temperatures (below 3 GK), the calculation of the rate has to rely on the
calculated cross section between the threshold at 5.1 MeV and the lowest data
point at 6.2 MeV. Because of the excellent reproduction of the energy
dependence of the \ran\ cross section we estimate an overall uncertainty of
less than 30\% for all temperatures.

This 30\% total uncertainty was estimated in the following way. The
experimental data have uncertainties of about 10\%. Thus, above T = 3 GK, the
rate is fully determined by the new experimental data, and accordingly the
uncertainty of the rate is of the order of 10\%. For the rates at lower
temperatures, the experimental cross sections have to be extrapolated towards
lower energies. The overall uncertainty of an ATOMKI-V2 prediction is $\leq$
factor 2 \citep{Mohr2020a}, and the uncertainty in the present case is reduced
to $\leq$ 50\% by the normalization of the ATOMKI-V2 calculation to the
experimental data. As even at the lowest relevant temperatures a significant
part of the Gamow window is covered by experimental data, the uncertainty of
the rate should not exceed 30\% at low temperatures. For simplicity, we have
used this 30\% as overall uncertainty at all temperatures which is a very
careful estimate. As will be shown in the following section, such an overall
uncertainty of the rate of 30\% is sufficient to constrain the nucleosynthesis
path in the weak r-process.

The obtained reaction rates are listed in Table \ref{tab:rate}. Previously
recommended rates in the widely used databases were based either on the
McFadden/Satchler AOMP, e.g.\ in REACLIB \citep{REACLIB,Cyburt2010a} and from
NON-SMOKER \citep{Rauscher2000a}, or on the Demetriou AOMP, e.g.\ in STARLIB
\citep{STARLIB,Sallaska2013a}. The predictions for the \zrvi \ran \moix\ rate
in the available databases vary by more than one order of magnitude, thus
leading to significant uncertainties in nucleosynthesis calculations. The
present recommended rate has a significantly reduced uncertainty of about 30\%
which allows stronger conclusions on the nucleosynthesis in the weak
r-process.
\begin{table}
\center
\caption{\label{tab:rate}Recommended astrophysical reaction rate \Nsv\ of the
  \zrvi \ran \moix\ reaction.
}
\begin{tabular}{ccccc}
\hline
\multicolumn{1}{c}{$T_9$} &
\multicolumn{1}{c}{\Nsv\ (cm$^3$\,s$^{-1}$\,mole$^{-1}$)} \\
\hline
  1.0  & 2.09$\times 10^{-25}$ \\
  1.5  & 1.36$\times 10^{-16}$ \\
  2.0  & 5.44$\times 10^{-12}$ \\
  2.5  & 5.01$\times 10^{-09}$ \\
  3.0  & 7.11$\times 10^{-07}$ \\
  4.0  & 7.03$\times 10^{-04}$ \\
  5.0  & 4.44$\times 10^{-02}$ \\
\hline
\end{tabular}
\end{table}

\section{Impact on weak r-process}
\label{sec:rproc}
We investigate the impact of the new experimental data on the nucleosynthesis of
lighter heavy elements in neutron-rich supernova ejecta. We use astrophysical
trajectories based on the neutrino-driven wind model of \cite{Bliss2018a}.
Each trajectory corresponds to a combination of astrophysical parameters which are
expected for neutrino-driven winds. The 36 trajectories under consideration (see Table I of
\cite{Bliss2020a}) cover electron fractions between 0.40 and 0.49,  entropies between
32 and 175 $k_{\rm{B}}$ per nucleon, and expansion timescales from 9.7 to 63.8 ms.
In that work, the authors identified the conditions for which ($\alpha$,n)
reactions have a significant impact on the final abundances.
Under such conditions, \cite{Bliss2020a} used 36 trajectories to identify key
($\alpha$,n) reactions.
The reaction $^{96}$Zr($\alpha$,n)$^{99}$Mo is in their list of key
reactions.
Our nucleosynthesis calculations are performed with the WinNet reaction network \citep{Winteler2012a}.
Reaction rates are taken from the JINA REACLIBV2.0 \citep{REACLIB,Cyburt2010a} library
except for ($\alpha$,n) reactions for which TALYS 1.6 with the global
$\alpha$OMP (GAOP) was used. The GAOP is based on \cite{Watanabe1958a}; for
more details see \cite{Pereira2016a,Bliss2018a,Bliss2020a}.
Replacing the $^{96}$Zr($\alpha$,n)$^{99}$Mo reaction rate with the values from
Tab.~\ref{tab:rate} results in a reduction of the final abundances above $Z = 44$.
For 17 of the 36 trajectories the largest reduction is higher than 10\% and for 6 of
them it is higher than 20\%.
More importantly, the reduced reaction-rate uncertainty leads to a significant
improvement in the accuracy of the nucleosynthesis predictions.
Following \cite{Bliss2020a} we estimate the uncertainty of the
$^{96}$Zr($\alpha$,n)$^{99}$Mo reaction rate calculated with the GAOP with the
factors 0.1 and 10 and the uncertainty of the present reaction rate with 30\%
(see Sect.~\ref{sec:res}).
\begin{figure}[ht]
\begin{center}
\includegraphics[width=0.99\linewidth]{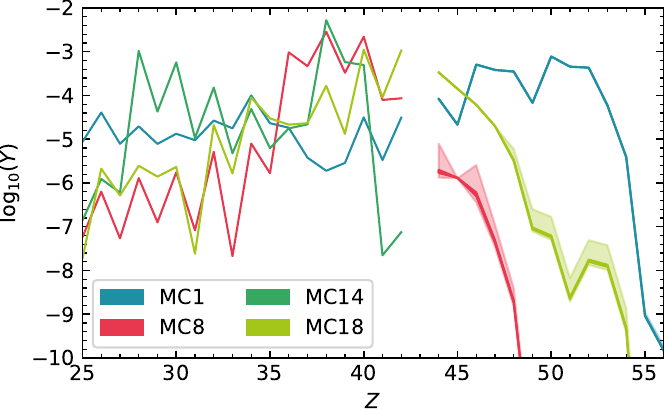}%
\end{center}
\caption{Elemental abundances for four trajectories from \cite{Bliss2020a}.
  The lightly shaded regions correspond to the uncertainties due to variations
of the previously used $^{96}$Zr($\alpha$,n)$^{99}$Mo reaction rate by factors
10 and 0.1.
The solid bands correspond to the uncertainties due to variations of the new
recommended reaction rate by 30\%.}
\label{fig:abund}
\end{figure}

\begin{figure}[ht]
\begin{center}
\includegraphics[width=0.99\linewidth]{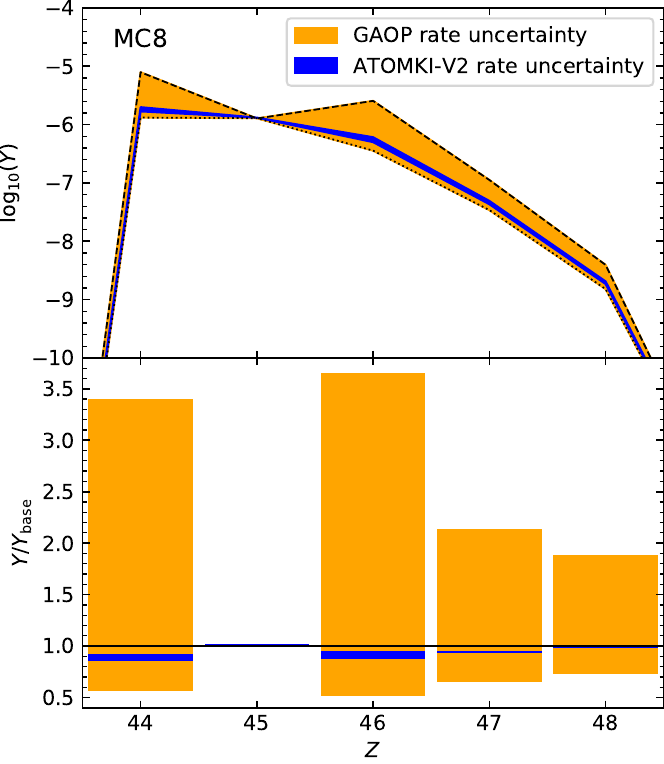}%
\end{center}
\caption{Influence of the $^{96}$Zr($\alpha$,n)$^{99}$Mo rate on trajectory MC8.
         Upper panel: Abundance uncertainty of elements between Ru and
         Cd. The orange and blue bands correspond to the previously used
         (GAOP) and the present reaction rate, respectively.
         The dashed and dotted lines show the abundance pattern calculated with
         upper and lower uncertainty estimation of the GAOP rate, respectively.
         Lower panel: Abundance uncertainties relative to the unvaried GAOP
         reaction rate,~$Y_{\rm base}$,
         in a linear scale.}
\label{fig:uncertaintiyMC8}
\end{figure}

In Fig.~\ref{fig:abund}, we present the impact of the reduced uncertainty of the
new experimentally based reaction rate for four representative trajectories from
\cite{Bliss2020a}.
Changes of the final abundances resulting from the variation of the GAOP and
ATOMKI-V2 $^{96}$Zr($\alpha$,n)$^{99}$Mo reaction rate are represented by the shaded
and solid bands, respectively (note that the figure shows solid colored bands
and not thick lines).
If large amounts of elements heavier than Tc are produced (e.g.,
trajectory MC1 in Fig.~\ref{fig:abund}), the abundances are not sensitive to
$^{96}$Zr($\alpha$,n)$^{99}$Mo, because the nucleosynthesis path runs along
more neutron-rich nuclei.
Trajectories that do not produce any elements beyond Mo (e.g., trajectory
MC14 in Fig.~\ref{fig:abund}) are not sensitive either.
For roughly half of the 36 trajectories, the variation of the previously used
$^{96}$Zr($\alpha$,n)$^{99}$Mo reaction rate leads to a significant spread (up
to a factor of 6 between the lower and upper estimate) in the elemental
abundances between Ru and Xe (e.g., trajectories MC8 and MC18 in
Fig.~\ref{fig:abund}).
In all of these trajectories, the lower uncertainty of the present reaction rate
leads to greatly improved accuracy in the final abundances.
In Fig.~\ref{fig:uncertaintiyMC8}, we show the abundances for trajectory MC8 in detail.
The orange and blue bands represent the uncertainty as estimated for the GAOP
and the ATOMKI-V2 reaction rate, respectively.
The dashed and dotted lines in the upper panel show the abundance pattern
calculated with upper and lower uncertainty estimation of the GAOP reaction
rate, respectively.
In the bottom panel, we show the uncertainty for each element relative to the
abundances calculated with the unvaried GAOP reaction rate,~$Y_{\rm base}$.
Since $^{96}$Zr($\alpha$,n)$^{99}$Mo forms a bottleneck for this trajectory, an
increase of the reaction rate results in higher abundances of elements heavier than Tc.
The ATOMKI-V2 reaction rate is slightly lower than the GAOP reaction rate and thus
the abundances are slightly lower than $Y_{\rm base}$.
An exception is the abundance of Rhodium which is not sensible to
$^{96}$Zr($\alpha$,n)$^{99}$Mo.
Rhodium possesses only one stable isotope, $^{103}$Rh, which in all trajectories
is mainly produced by the decay of $^{103}$Nb.
Its abundance is therefore not correlated to $^{96}$Zr($\alpha$,n)$^{99}$Mo.

In summary, the reduction of the uncertainty to 30\% is sufficient to get
very accurate abundances.
This accuracy is crucial for comparing theoretical nucleosynthesis calculations
with observations.
A similar reduction of the uncertainties for other reactions is necessary to
reliably compare nucleosynthesis calculations with observations.
The PBTM should allow for such a reduction of uncertainties; a detailed investigation is in preparation.
This will allow to constrain the astrophysical site of the weak r-process and to
further understand core-collapse supernovae and the origin of the lighter heavy
elements.

\section{Summary and Conclusions}
\label{sec:sum}
In a recent sensitivity study of the weak r-process \citep{Bliss2020a}, the $^{96}$Zr($\alpha$,n)$^{99}$Mo reaction was identified as a bottleneck for the nucleosynthesis between ruthenium and cadmium, i.e.\ for nuclei with $Z = 44 - 48$. The typically assumed uncertainties of \ran\ reaction rates of a factor of 10 lead to significant uncertainties for the nucleosynthsis yields in the weak r-process of about a factor of 5-6; thus the nuclear uncertainties prevent any robust astrophysical conclusion.

In the present study, the cross section of the $^{96}$Zr($\alpha$,n)$^{99}$Mo
reaction has been measured for the first time from energies slightly above the
reaction threshold at 5.1 MeV up to about 12.5 MeV, thus covering the region
relevant of temperatures relevant for the weak r-process. The chosen
activation technique provides the total production cross section of $^{99}$Mo
which is an excellent basis for the calculation of the astrophysical
production rate of molybdenum from $^{96}$Zr by $\alpha$-induced
reactions. The high precision experimental data have been analyzed in the
statistical model, using global $\alpha$OMP's and complemented by the recently
suggested barrier transmission model. It was found that the new approach ---
re-scaled by 0.65 --- excellently reproduces the new experimental data. The
scaled best-fit was used to calculate the astrophysical reaction rates as a
function of temperature. For the full temperature range of the weak r-process,
the uncertainty of the reaction rate could be drastically reduced from the
usually assumed factor of 10 down to about 30\%.

A repetition of the nucleosynthesis calculations of \cite{Bliss2020a} with the
new experimentally based $^{96}$Zr($\alpha$,n)$^{99}$Mo reaction rate and its
small uncertainties leads to very well-constrained nucleosynthesis yields for
the $Z = 44 - 48$ range. As the barrier transmission model, implemented as
ATOMKI-V2 potential, is typically able to predict $\alpha$-induced reaction
cross sections with uncertainties below a factor of two, a re-calculation of
the full weak r-process network with updated rates from the barrier
transmission model will lead to more robust nucleosynthesis yields which in
turn should enable a major step towards stringent constraints for the
astrophysical conditions and the site of the weak r-process.

\acknowledgements{
The authors thank Julia Bliss, Fernando Montes, Jorge Pereira, and Zs.\ F\"ul\"op for valuable discussions. This work was supported by NKFIH (NN128072, K120666, K134197), and by the \'UNKP-20-5-DE-2 New National Excellence Program of the Ministry of Human Capacities of Hungary. G. G. Kiss acknowledges support from the J\'anos Bolyai research fellowship of the Hungarian Academy of Sciences. MJ and AA were supported by the ERC Starting Grant EUROPIUM-677912, Deutsche Forschungsgemeinschaft through SFB 1245, and Helmholtz Forschungsakademie Hessen für FAIR. This work has benefited from the COST Action “ChETEC” (CA16117) supported by COST (European Cooperation in Science and Technology). The nucleosynthesis computations were performed on the Lichtenberg High Performance Computer (TU Darmstadt).}

\bibliographystyle{apj}
\bibliography{main}

\end{document}